\documentclass[12pt]{article}
\usepackage[latin9]{inputenc}
\usepackage{amsmath}
\usepackage{amssymb}
\usepackage{esint}
\usepackage[unicode=true,pdfusetitle,
 bookmarks=true,bookmarksnumbered=false,bookmarksopen=false,
 breaklinks=false,pdfborder={0 0 0},pdfborderstyle={},backref=false,colorlinks=false]
 {hyperref}
\usepackage{breakurl}

\makeatletter
\numberwithin{equation}{section}

\usepackage{amsfonts}
\usepackage{ytableau} % Young tableaux
\usepackage[square,numbers,sort&compress]{natbib}

\makeatother

\begin{document}
\begin{flushright}
FIAN/TD/2019-6\\
 
\par\end{flushright}

\vspace{0.5cm}
 
\begin{center}
\textbf{\large{}On unfolded off-shell formulation for higher-spin
theory}{\large\par}
\par\end{center}

\begin{center}
\vspace{0.1cm}
\textbf{N.G.~Misuna}\\
 \vspace{0.5cm}
 \textit{I.E. Tamm Department of Theoretical Physics, Lebedev Physical
Institute,}\\
 \textit{ Leninsky prospect 53, 119991, Moscow, Russia}\\
\par\end{center}

\begin{center}
\textit{Moscow Institute of Physics and Technology,}\\
 \textit{ Institutsky pereulok 9, 141701, Dolgoprudny, Moscow region,
Russia}\\
\par\end{center}

\begin{center}
\vspace{0.6cm}
misuna@phystech.edu \\
\par\end{center}

\vspace{0.4cm}

\begin{abstract}
\noindent We present an unfolded off-shell formulation for free massless
higher-spin fields in $4d$ Minkowski space in terms of spinorial
variables. This system arises from the on-shell one by the addition
of external higher-spin currents, for which we find an unfolded description.
Also we show that this off-shell system can be interpreted as Schwinger\textendash Dyson
equations and restore two-point functions of higher-spin fields this
way.
\end{abstract}

\section{Introduction}

Higher-spin (HS) gauge theory (for a review see \emph{e.g. }\cite{Vas_StarProduct,DidSkv_Elements}),
describing interactions of massless fields of all spins and possessing
infinite gauge symmetry, is of great interest for high-energy physics.
Firstly, it is regarded as possible symmetric high-energy phase of
string theory \cite{Gross}, secondly, it provides an example of weak-weak
$AdS/CFT$ correspondence, being dual to different conformal vectorial
models according to Klebanov\textendash Polyakov conjecture \cite{KlebPol}.
Available formulation of HS theory represents a full nonlinear system
of classical equations of motion written in the so-called unfolded
form \textendash{} Vasiliev equations \cite{Vas_Eq,Vas_More}. Unfolded
dynamics approach is a first-order formalism, possessing a manifest
coordinate-independence (due to exterior forms language) and allowing
an easy control of gauge symmetries of the theory. However, obtaining
conventional field-theoretical information from the unfolded formulation
turns out to be a very nontrivial problem. Up to date only cubic HS
vertices were managed to be extracted from Vasiliev equations and
compared with $AdS/CFT$ predictions \cite{GioYin_3pt,GioYin_Twist,DidVas_3pt,SezgSkvZhu,GelfVas_1form,Misuna}.

One of the main problems of HS theory is the absence of a full nonlinear
action (see, however, \cite{BoulSun_Action}), that does not allow
the use of standard $AdS/CFT$ techniques and the study of quantization
issue. A general systematic method of classifying all gauge-invariant
functionals of unfolded system was proposed in \cite{Vas_Action}.
It consists in studying cohomologies of certain operator determined
by unfolded equations of the theory. In the case of an on-shell unfolded
system (\emph{i.e.} when it contains dynamical equations on primary
fields) such functionals correspond to conserved charges. In the case
of an off-shell unfolded system (when unfolded equations represent
a set of constraints expressing descendants via primaries with prescribed
gauge symmetries) these functionals can be considered as potential
actions.

Thus, in order to proceed in the search for HS action, one should
find an off-shell completion of Vasiliev equations \textendash{} an
unfolded system with the same spectrum of primary fields, proper gauge
transformations laws, but without any dynamical equations. Such off-shell
construction, resulting from relaxing tracelesness condition, for
Vasiliev theory of totally symmetric bosonic HS fields in arbitrary
dimension \cite{Vas_AdsD} was presented in \cite{SagnSezgSund}.
However, the most elaborated is another version of Vasiliev system,
namely, $4d$ theory formulated in terms of spinors, which in most
cases is the only one where practical computations are feasible. Here
traces are absent by construction, so the method of \cite{SagnSezgSund}
is inapplicable. In this paper we propose an alternative way to look
for off-shell completion for this theory, by coupling it to external
currents. Concretely, we present an off-shell completion of unfolded
spinorial system of free HS in $4d$ flat spacetime. One of the advantage
of the proposed approach is that it allows one to look for quantum
correlation functions via interpreting off-shell equations as the
Schwinger\textendash Dyson ones. We illustrate this by finding two-point
functions of HS fields from the off-shell system we built.

The paper is organized as follows. In Section \ref{Sec2} we recall
some facts about unfolded dynamics approach and HS equations. In Section
\ref{Sec3} we construct unfolded description of Fronsdal currents
and use it to find an off-shell completion of free HS equations. In
\ref{Sec4} two-point correlation functions of HS fields are evaluated
from the off-shell system of Section \ref{Sec3}. In \ref{Conclusion}
we present our conclusions.

\section{Unfolded formulation and free on-shell HS equations\label{Sec2}}

Unfolding of the theory means reformulating it via equations of the
form
\begin{equation}
\mathrm{d}W^{A}\left(x\right)=G^{A}\left(W\right),\label{unf_eq}
\end{equation}
where $\mathrm{d}$ is spacetime de Rham differential and $W^{A}\left(x\right)$
are differential forms representing unfolded fields with $A$ denoting
all indices they carry. The identity $\mathrm{d}^{2}\equiv0$ imposes
the following consistency condition on the system \eqref{unf_eq}
\begin{equation}
G^{B}\dfrac{\delta G^{A}}{\delta W^{B}}\equiv0.\label{unf_consist}
\end{equation}
A consistent unfolded system is manifestly invariant under a set of
gauge transformations
\begin{equation}
\delta W^{A}=\mathrm{d}\varepsilon^{A}\left(x\right)-\varepsilon^{B}\dfrac{\delta G^{A}\left(W\right)}{\delta W^{B}}.\label{unf_gauge_transf}
\end{equation}

A simple example of unfolded system is provided by equations describing
Minkowski space. In this case unfolded fields are 1-forms of vielbein
$e^{a}=e^{a}{}_{\underline{m}}\mathrm{d}x^{\underline{m}}$ and Lorentz
spin-connection $\omega_{L}^{a,b}=\omega_{L}^{a,b}{}_{\underline{m}}\mathrm{d}x^{\underline{m}}=-\omega_{L}^{b,a}$
which obey
\begin{eqnarray}
 &  & \mathrm{d}e^{a}+\omega_{L}^{a,b}e_{b}=0,\\
 &  & \mathrm{d}\omega_{L}^{a,b}+\omega_{L}^{a,}{}_{c}\omega_{L}^{c,b}e_{b}=0.
\end{eqnarray}

Another example is an unfolded description of free massless scalar
field. Appropriate system look as follows
\begin{equation}
D^{L}C_{a\left(n\right)}=e^{b}C_{ba\left(n\right)},\label{unf_scalar}
\end{equation}
where $D^{L}=\mathrm{d}+\omega_{L}$ is Lorentz-covariant derivative
and $C_{a\left(n\right)}\left(x\right)$ are symmetric rank-$n$ Lorentz
tensors. In Cartesian coordinates
\begin{equation}
e^{a}{}_{\underline{m}}=\delta^{a}{}_{\underline{m}},\qquad\omega_{L}^{a,b}=0\label{Cartes_coord}
\end{equation}
one can solve \eqref{unf_scalar} as
\begin{equation}
C_{a\left(n\right)}\left(x\right)=\partial_{a_{1}}...\partial_{a_{n}}C\left(x\right),\label{scalar_solut}
\end{equation}
which shows that $C_{a\left(n\right)}\left(x\right)$, $n>0$ are
descendant fields, forming the tower of all derivatives of the primary
scalar $C\left(x\right)$. If there is no further constraints on the
unfolded fields, the system \eqref{unf_scalar} is off shell in the
sense that there is no any differential constraints on the primary
field $C\left(x\right)$. However if one requires $C_{a\left(n\right)}\left(x\right)$
to be traceless, then from \eqref{scalar_solut} Klein-Gordon equation
$\square C\left(x\right)=0$ follows. So the same unfolded system
\eqref{unf_scalar} for traceless tensors describes on-shell scalar
field.

A spectrum of unfolded fields of HS theory includes master 1-form
$\omega$ and master 0-form $C$ that depend on a pair of auxiliary
Lorentz vectors $Y_{1}^{a}$ and $Y_{2}^{a}$
\begin{equation}
\omega\left(Y|x\right)=\sum_{n\geq m}\omega_{a\left(n\right),b\left(m\right)}Y_{1}^{a_{1}}...Y_{1}^{a_{n}}Y_{2}^{b_{1}}...Y_{2}^{b_{m}},\qquad C\left(Y|x\right)=\sum_{n\geq m}C_{a\left(n\right),b\left(m\right)}Y_{1}^{a_{1}}...Y_{1}^{a_{n}}Y_{2}^{b_{1}}...Y_{2}^{b_{m}}.
\end{equation}
where tensors possess symmetry of two-row Young diagrams. Submodule
describing spin-$s$ field is formed by $\omega_{a\left(s-1\right),b\left(t\right)}$
(Fronsdal gauge spin-$s$ field and its first $\left(s-1\right)$
derivatives) and $C_{a\left(s+t\right),b\left(s\right)}$ (gauge-invariant
HS curvatures and infinite towers of their descendants). To write
down on-shell equations, one requires all tensors to be traceless.

The most elaborated is $4d$ Vasiliev system due to the isomorphism
$so\left(3,1\right)\approx sl\left(2,\mathbb{C}\right)$ that allows
one to replace auxiliary vectors $Y_{i}^{a}$ with auxiliary spinor
variables $Y^{A}=\left(y^{\alpha},\bar{y}^{\dot{\alpha}}\right)$,
$\alpha,\dot{\alpha}=\overline{1,2}$. In this case requirement of
tracelessness for Lorentz tensors gets replaced with a simple condition
of symmetricity of $Y^{A}$. $4d$ unfolded master-fields now are
\begin{equation}
\omega\left(Y|x\right)=\sum_{n,m}\dfrac{1}{n!m!}\omega_{\alpha\left(n\right),\dot{\beta}\left(m\right)}y^{\alpha_{1}}...y^{\alpha_{n}}\bar{y}^{\dot{\beta}_{1}}...\bar{y}^{\dot{\beta}_{m}},\quad C\left(Y|x\right)=\sum_{n,m}\dfrac{1}{n!m!}C_{\alpha\left(n\right),\dot{\beta}\left(m\right)}y^{\alpha_{1}}...y^{\alpha_{n}}\bar{y}^{\dot{\beta}_{1}}...\bar{y}^{\dot{\beta}_{m}}.
\end{equation}
In terms of them one can write down an unfolded system describing
propagation of free massless HS fields (so-called Central On-mass-shell
Theorem) \cite{Vas_More}. For Minkowski background it is
\begin{eqnarray}
 &  & D^{L}\omega\left(Y|x\right)+e^{\alpha\dot{\beta}}y_{\alpha}\bar{\partial}_{\dot{\beta}}\Pi^{-}\omega\left(Y|x\right)+e^{\alpha\dot{\beta}}\partial_{\alpha}\bar{y}_{\dot{\beta}}\Pi^{+}\omega\left(Y|x\right)=\nonumber \\
 &  & =\dfrac{i}{4}\bar{H}^{\dot{\alpha}\dot{\beta}}\bar{\partial}_{\dot{\alpha}}\bar{\partial}_{\dot{\beta}}C\left(0,\bar{y}|x\right)+\dfrac{i}{4}H^{\alpha\beta}\partial_{\alpha}\partial_{\beta}C\left(y,0|x\right),\label{OnShTh_flat_1}\\
 &  & D^{L}C\left(Y|x\right)+ie^{\alpha\dot{\beta}}\partial_{\alpha}\bar{\partial}_{\dot{\beta}}C\left(Y|x\right)=0.\label{OnShTh_flat_2}
\end{eqnarray}
Here $H^{\alpha\beta}=e^{\alpha}{}_{\dot{\gamma}}e^{\beta\dot{\gamma}}$,
$\bar{H}^{\dot{\alpha}\dot{\beta}}=e_{\gamma}\text{}^{\dot{\alpha}}e^{\gamma\dot{\beta}}$
and we introduced projectors $\Pi^{+}$ ($\Pi^{-}$) to components
with $N>\bar{N}$ ($N<\bar{N}$, respectively), where $N=y^{\alpha}\partial_{\alpha}$
and $\bar{N}=\bar{y}^{\dot{\alpha}}\bar{\partial}_{\dot{\alpha}}$
count the number of $y$ and $\bar{y}$. The system \eqref{OnShTh_flat_1}-\eqref{OnShTh_flat_2}
splits into independent subsystems with $\left(N+\bar{N}\right)\omega=\left(2s-2\right)\omega$,
$|N-\bar{N}|C=2sC$ that describe spin-$s$ field. Our goal is to
build and off-shell completion of \eqref{OnShTh_flat_1}-\eqref{OnShTh_flat_2}.

\section{HS sources and off-shell extension for HS equations\label{Sec3}}

As was mentioned, for Vasiliev system formulated in terms of Lorentz
tensors one can formulate an off-shell generalization via relaxing
the tracelessness projections \cite{SagnSezgSund}. Another example
of off-shell HS system is presented in \cite{Vas_Action}, where it
is shown that unfolded system consisting of covariant flatness and
covariant constancy equations for 1- and 0-forms taking values in
$d$-dimensional oscillator algebra can be treated for special vacuum
solution as nonlinear off-shell HS theory in Minkowski space. Corresponding
HS fields however turn out to be traceful, differing from standard
Fronsdal fields. How to perform proper reduction of this system that
would drive out traces remains unclear.

The situation with $4d$ spinorial formulation of Vasiliev system
is peculiar, because in this case it is the commutativity of auxiliary
spinors $Y^{A}$ that puts the system on-shell and there is no trace
projections that could be relaxed. To proceed in this case we choose
a following strategy. Consider a massless scalar field obeying Klein-Gordon
equation $\square\phi\left(x\right)=0$. This system is on shell.
Now deform it writing $\square\phi\left(x\right)=J\left(x\right)$.
If the source $J\left(x\right)$ is some given function, say, field-dependent
correction describing interactions or fixed external field, then the
theory remains on shell. However if we treat $\square\phi\left(x\right)=J\left(x\right)$
as equation determining $J\left(x\right)$, then our theory becomes
off-shell one. Now it contains two scalar fields, primary $\phi\left(x\right)$
and its descendant $J\left(x\right)$, without any differential restrictions
on $\phi\left(x\right)$.

Thus we can build an off-shell completion for \eqref{OnShTh_flat_1}-\eqref{OnShTh_flat_2}
by coupling it to sources that obeys no constraints other than required
by consistency condition. In standard formulation Fronsdal equation
for double-traceless spin-$s$ field $\phi_{a\left(s\right)}\left(x\right)$
coupled to double-traceless external current $J_{a\left(s\right)}\left(x\right)$
has the form
\begin{equation}
\square\phi_{a\left(s\right)}-s\partial_{a}\partial^{b}\phi_{ba\left(s-1\right)}+\frac{s\left(s-1\right)}{2}\partial_{a}\partial_{a}\phi^{b}\mbox{}_{ba\left(s-2\right)}=J_{a\left(s\right)},
\end{equation}
with consistency condition for the current
\begin{equation}
\partial^{b}J_{ba\left(s-1\right)}=\frac{\left(s-1\right)}{2}\partial_{a}J^{b}\mbox{}_{ba\left(s-2\right)}.\label{HS_conserv}
\end{equation}

First let us consider an example of off-shell completion for unfolded
scalar field. This field is described on-shell by a subsystem of \eqref{OnShTh_flat_2}
with $NC\left(Y|x\right)=\bar{N}C\left(Y|x\right)$
\begin{equation}
D^{L}C+ie^{\alpha\dot{\beta}}\partial_{\alpha}\bar{\partial}_{\dot{\beta}}C=0.\label{scalar_onsh}
\end{equation}
(which in fact is \eqref{unf_eq} rewritten in terms of spinors).
In this case the source is just another (unconstrained) scalar field.
We can describe it by the 0-form master-field $J\left(Y|x\right)$,
$NJ=\bar{N}J$ and put it to the r.h.s. of \eqref{scalar_onsh}. Next
we should write down an unfolded system for $J\left(Y|x\right)$.
To this end we can use the same equation \eqref{scalar_onsh} with
$C$ replaced with $J$, but now to avoid imposing $\square J=0$
we once again should introduce one more ``source for source'' to
its r.h.s. Obviously this process is infinitely repeating. So it is
convenient to introduce a new parameter $b$ and organize the whole
sequence of sources as expansion in it
\begin{equation}
J\left(Y|b|x\right)=\sum_{k=0}^{\infty}\frac{b^{k}}{k!}J^{\left(k\right)}\left(Y|x\right).\label{b_expansion}
\end{equation}

Then an unfolded system describing off-shell scalar field takes the
form
\begin{eqnarray}
 &  & D^{L}C+ie^{\alpha\dot{\beta}}\partial_{\alpha}\bar{\partial}_{\dot{\beta}}C=ie^{\alpha\dot{\beta}}y_{\alpha}\bar{y}_{\dot{\beta}}\frac{1}{\left(N+1\right)\left(N+2\right)}J\left(b=0\right),\label{scal_off_1}\\
 &  & D^{L}J+ie^{\alpha\dot{\beta}}\partial_{\alpha}\bar{\partial}_{\dot{\beta}}J=ie^{\alpha\dot{\beta}}y_{\alpha}\bar{y}_{\dot{\beta}}\frac{1}{\left(N+1\right)\left(N+2\right)}\frac{\partial}{\partial b}J,\label{scal_off_2}
\end{eqnarray}
with condition $NC=\bar{N}C$, $NJ=\bar{N}J$. $N$-dependent coefficients
in \eqref{scal_off_1}-\eqref{scal_off_2} are fixed by consistency
condition \eqref{unf_consist}. By analyzing this system in Cartesian
coordinates it is easy to see that $b$-expansion \eqref{b_expansion}
is in fact an expansion in boxes of off-shell primary scalar field
$C\left(x\right)$:\textbf{
\begin{equation}
J_{\alpha\left(n\right),\dot{\alpha}\left(n\right)}^{\left(k\right)}\left(x\right)\sim\left(\frac{\partial}{\partial x^{\alpha\dot{\alpha}}}\right)^{n}\square^{k+1}C\left(x\right).
\end{equation}
}

Let us note, that this construction also proposes a simple way of
imposing higher-order equations. For instance, to get an unfolded
description of scalar field subjected to equation
\begin{equation}
\square^{n}C\left(x\right)=0,
\end{equation}
 one should simply restrict the limit of summation in \eqref{b_expansion}
to $\left(n-1\right)$.

For HS sources the situation is more complicated, because possible
tensor symmetries of descendants are richer \textendash{} we can antisymmetrize
derivatives with spin indices of the primary source field or take
divergences of it. Moreover, HS sources are double-traceless and obey
the generalized conservation law \eqref{HS_conserv}. These two properties
can be treated as follows: spin-$s$ source represents a combination
of two traceless tensors of rank $s$ and $\left(s-2\right)$ with
the only condition that rank-$s$ tensor has a divergence proportional
to the first traceless symmetrized derivative of the rank-$\left(s-2\right)$
one; apart from that tensors are unrestricted. So it is easier to
start with considering the following problem: what is the unfolded
description of unconstrained traceless rank-$n$ Lorentz tensor field
$T_{a\left(n\right)}\left(x\right)$.

As the first step consider the case of conserved field obeying Klein-Gordon
equation
\begin{equation}
\square T_{a\left(n\right)}=0,\qquad\partial^{b}T_{ba\left(n-1\right)}=0.\label{boxT_divT}
\end{equation}
Let us analyze the spectrum of descendants it generates. In the language
of Young diagrams, we start with the one-row diagram of length $n$
and then successively add one cell step by step (successively differentiate
primary $T_{a\left(n\right)}$). Then, taking into account that all
derivatives are automatically symmetrized, we see that the space of
descendants gets parameterized by the set of all one- and two-row
traceless Young diagrams with the first row of no less than $n$ cells
and the second row (if presented) with no more than $n$ cells. In
the language of multispinors, this is equivalent to the set of $T_{\alpha\left(n\right),\dot{\beta}\left(m\right)}\left(x\right)$
with $n+m\geq2s$, $|n-m|\leq2s$. (Let us remind that traceless tensor
$T_{a\left(n\right),b\left(m\right)}$ with Young symmetry corresponds
to a pair of symmetric multispinors $T_{\alpha\left(n+m\right),\dot{\beta}\left(n-m\right)}$
and $\bar{T}_{\alpha\left(n-m\right),\dot{\beta}\left(n+m\right)}$).

Corresponding unfolded system must present some generalization of
\eqref{scalar_onsh}. An operator $ie^{\alpha\dot{\beta}}\partial_{\alpha}\bar{\partial}_{\dot{\beta}}$
corresponds to the adding cell to the upper (for scalar field \textendash{}
the only) row. This must be complemented by operators of the form
$e^{\alpha\dot{\beta}}y_{\alpha}\bar{\partial}_{\dot{\beta}}\Pi^{-}$
and $e^{\alpha\dot{\beta}}\partial_{\alpha}\bar{y}_{\dot{\beta}}\Pi^{+}$
which account for the possibility of adding cell to the bottom row.
Fixing the coefficients by consistency, one arrives at the following
unfolded system, corresponding to \eqref{boxT_divT},
\begin{equation}
D^{L}T+ie^{\alpha\dot{\beta}}\partial_{\alpha}\bar{\partial}_{\dot{\beta}}T+e^{\alpha\dot{\beta}}y_{\alpha}\bar{\partial}_{\dot{\beta}}\frac{1}{\left(N+1\right)\left(N+2\right)}\Pi^{-}T+e^{\alpha\dot{\beta}}\partial_{\alpha}\bar{y}_{\dot{\beta}}\frac{1}{\left(\bar{N}+1\right)\left(\bar{N}+2\right)}\Pi^{+}T=0.
\end{equation}

At the second step let us get rid of the constraint $\square T_{a\left(n\right)}=0$.
To this end, as in the scalar field example, we introduce a dependence
on the additional parameter $b$ (``power of boxes'') and add corresponding
$b$-dependent terms to the r.h.s. of the unfolded equations. As taking
box means removing one cell from the Young diagram of descendant,
these terms must be of the form $ie^{\alpha\dot{\beta}}y_{\alpha}\bar{y}_{\dot{\beta}}$(removing
cell from the upper row, presented already in the scalar case), $e^{\alpha\dot{\beta}}y_{\alpha}\bar{\partial}_{\dot{\beta}}\Pi^{+0}$
and $e^{\alpha\dot{\beta}}\partial_{\alpha}\bar{y}_{\dot{\beta}}\Pi^{-0}$
(removing from the bottom one), where $\Pi^{+0}$ ($\Pi^{-0}$) are
projectors to $N\geq\bar{N}$ ($N\leq\bar{N}$). After determining
coefficients from consistency condition one gets
\begin{eqnarray}
 &  & D^{L}T\left(Y|b|x\right)+ie^{\alpha\dot{\beta}}\partial_{\alpha}\bar{\partial}_{\dot{\beta}}T+e^{\alpha\dot{\beta}}y_{\alpha}\bar{\partial}_{\dot{\beta}}\frac{1}{\left(N+1\right)\left(N+2\right)}\Pi^{-}T+e^{\alpha\dot{\beta}}\partial_{\alpha}\bar{y}_{\dot{\beta}}\frac{1}{\left(\bar{N}+1\right)\left(\bar{N}+2\right)}\Pi^{+}T=\nonumber \\
 &  & =ie^{\alpha\dot{\beta}}y_{\alpha}\bar{y}_{\dot{\beta}}\frac{1}{\left(N+1\right)\left(N+2\right)\left(\bar{N}+1\right)\left(\bar{N}+2\right)}\frac{\partial}{\partial b}T-\nonumber \\
 &  & -e^{\alpha\dot{\beta}}y_{\alpha}\bar{\partial}_{\dot{\beta}}\frac{1}{\left(N+1\right)\left(N+2\right)}\Pi^{+0}\frac{\partial}{\partial b}T-e^{\alpha\dot{\beta}}\partial_{\alpha}\bar{y}_{\dot{\beta}}\frac{1}{\left(\bar{N}+1\right)\left(\bar{N}+2\right)}\Pi^{-0}\frac{\partial}{\partial b}T.
\end{eqnarray}
Let us note that this system is of interest by itself, because it
provides a description of traceless conserved tensor fields of arbitrary
ranks, \emph{i.e. $4d$} conformal HS currents, which are the subject
of the study in the literature \cite{ConfCurr1,ConfCurr2,ConfCurr3}.

The last step is to remove conservation condition $\partial^{b}T_{ba\left(n-1\right)}=0$.
This requires the introduction of one more parameter $f$ and corresponding
expansion (``power of divergences''), analogous to $b$ (of course,
an expansion in $f$ is finite, as one can take no more than $n$
divergences of rank-$n$ tensor)
\begin{equation}
T\left(Y|b,f|x\right)=\sum_{p=0}^{\infty}\sum_{q=0}^{n}\frac{f^{q}}{q!}\frac{b^{p}}{p!}T^{\left(p,q\right)}\left(Y|x\right).\label{f_expansion}
\end{equation}

From the standpoint of Young diagrams taking divergencies reveals
in the same way as taking box \textendash{} both remove one cell.
The difference however is that modules corresponding to higher powers
of $f$ (higher divergences) describes tensor fields of less ranks,
so for them one has $N+\bar{N}\geq2\left(n-f\frac{\partial}{\partial f}\right)$,
$|N-\bar{N}|\leq2\left(n-f\frac{\partial}{\partial f}\right)$. Thus
we have
\begin{eqnarray}
 &  & D^{L}T+ie^{\alpha\dot{\beta}}\partial_{\alpha}\bar{\partial}_{\dot{\beta}}T+e^{\alpha\dot{\beta}}y_{\alpha}\bar{\partial}_{\dot{\beta}}\frac{1}{\left(N+1\right)\left(N+2\right)}\Pi^{-}T+e^{\alpha\dot{\beta}}\partial_{\alpha}\bar{y}_{\dot{\beta}}\frac{1}{\left(\bar{N}+1\right)\left(\bar{N}+2\right)}\Pi^{+}T=\nonumber \\
 &  & =ie^{\alpha\dot{\beta}}y_{\alpha}\bar{y}_{\dot{\beta}}\frac{1}{\left(N+1\right)\left(N+2\right)\left(\bar{N}+1\right)\left(\bar{N}+2\right)}\left(\frac{\partial}{\partial b}+\frac{\partial}{\partial f}\right)T-\nonumber \\
 &  & -e^{\alpha\dot{\beta}}y_{\alpha}\bar{\partial}_{\dot{\beta}}\frac{1}{\left(N+1\right)\left(N+2\right)}\Pi^{+0}\left(\frac{\partial}{\partial b}+\frac{\partial}{\partial f}\right)T-e^{\alpha\dot{\beta}}\partial_{\alpha}\bar{y}_{\dot{\beta}}\frac{1}{\left(\bar{N}+1\right)\left(\bar{N}+2\right)}\Pi^{-0}\left(\frac{\partial}{\partial b}+\frac{\partial}{\partial f}\right)T.\nonumber \\
\label{T_eq}
\end{eqnarray}
This system solves our auxiliary problem, describing totally unconstrained
traceless rank-$n$ tensor field $T_{a\left(n\right)}\left(x\right)$
in terms of master 0-form $T\left(Y|b,f|x\right)$ such that $\left(N+\bar{N}\right)T\geq2\left(n-f\frac{\partial}{\partial f}\right)T$,
$|N-\bar{N}|T\leq2\left(n-f\frac{\partial}{\partial f}\right)T$.

Now, using this, we can write down a full system describing a space
of HS sources. To avoid degeneracy in spins we need to introduce last
additional parameter $m$ that encodes spin of the current
\begin{equation}
T\left(Y|b,f,m|x\right)=\sum_{2s-4=0}^{\infty}\sum_{p=0}^{\infty}\sum_{q=0}^{s-2}\frac{m^{2s}}{\left(2s\right)!}\frac{f^{q}}{q!}\frac{b^{p}}{p!}T^{\left(p,q,2s-4\right)}\left(Y|x\right).\label{T_fin_exp}
\end{equation}
This parameter allows one to distinguish between tensors of the same
Young symmetry but corresponding to different spins (\emph{e.g.} $T_{a\left(k\right)}$
may equally describe primary source of spin-$k$, or $k$-th derivative
of scalar source, or divergence of spin-$\left(k+1\right)$ source
\emph{etc.}) The role of parameters $b$, $f$ and $m$ (and need
for them) can be illustrated by the following relation, giving a manifest
expression for particular unfolded descendant in terms of derivatives
of primary field:
\begin{equation}
T^{\left(p,q,2s-4\right)}{}_{\alpha\left(n\right),\dot{\alpha}\left(n\right)}\left(x\right)\sim\square^{p}\left(\partial^{b}\right)^{q}\left(\partial_{a}\right)^{n-s+2+q}T_{b\left(q\right)a(s-2-q)}\left(x\right).
\end{equation}

To describe Fronsdal currents, we treat $T$ as a space of totally
unconstrained traces of currents (this is why we started sum with
$s=2$ in \eqref{T_fin_exp}) and complement it with analogous system
for $J\left(Y|b,f,m|x\right)$
\begin{equation}
J\left(Y|b,f,m|x\right)=\sum_{s=0}^{\infty}\sum_{p=0}^{\infty}\sum_{q=0}^{s}\frac{m^{2s}}{\left(2s\right)!}\frac{f^{q}}{q!}\frac{b^{p}}{p!}J^{\left(p,q,2s\right)}\left(Y|x\right)\label{J_fin_exp}
\end{equation}
\textendash{} traceless components of Fronsdal currents, that have
divergences proportional to first symmetrized derivatives of $T$:
\begin{eqnarray}
 &  & D^{L}J+ie^{\alpha\dot{\beta}}\partial_{\alpha}\bar{\partial}_{\dot{\beta}}J+e^{\alpha\dot{\beta}}y_{\alpha}\bar{\partial}_{\dot{\beta}}\frac{1}{\left(N+1\right)\left(N+2\right)}\Pi^{-}J+e^{\alpha\dot{\beta}}\partial_{\alpha}\bar{y}_{\dot{\beta}}\frac{1}{\left(\bar{N}+1\right)\left(\bar{N}+2\right)}\Pi^{+}J=\nonumber \\
 &  & =ie^{\alpha\dot{\beta}}y_{\alpha}\bar{y}_{\dot{\beta}}\frac{1}{\left(N+1\right)\left(N+2\right)\left(\bar{N}+1\right)\left(\bar{N}+2\right)}\left(\frac{\partial}{\partial b}J+T\right)-\nonumber \\
 &  & -e^{\alpha\dot{\beta}}y_{\alpha}\bar{\partial}_{\dot{\beta}}\frac{1}{\left(N+1\right)\left(N+2\right)}\Pi^{+0}\left(\frac{\partial}{\partial b}J+T\right)-e^{\alpha\dot{\beta}}\partial_{\alpha}\bar{y}_{\dot{\beta}}\frac{1}{\left(\bar{N}+1\right)\left(\bar{N}+2\right)}\Pi^{-0}\left(\frac{\partial}{\partial b}J+T\right).\nonumber \\
\label{J_eq}
\end{eqnarray}
Conditions on $Y$-powers for $J$ and $T$ now can be formulated
as follows
\begin{eqnarray}
 &  & \left(N+\bar{N}\right)J\geq\left(m\frac{\partial}{\partial m}-2f\frac{\partial}{\partial f}\right)J,\quad|N-\bar{N}|J\leq\left(m\frac{\partial}{\partial m}-2f\frac{\partial}{\partial f}\right)J,\\
 &  & \left(N+\bar{N}\right)T\geq\left(m\frac{\partial}{\partial m}-4-2f\frac{\partial}{\partial f}\right)T,\quad|N-\bar{N}|T\leq\left(m\frac{\partial}{\partial m}-4-2f\frac{\partial}{\partial f}\right)T.
\end{eqnarray}

Finally, to get an off-shell completion of unfolded HS system \eqref{OnShTh_flat_1}-\eqref{OnShTh_flat_2}
we must couple it to unfolded HS currents \eqref{T_eq}, \eqref{J_eq}
we found. The result is
\begin{eqnarray}
 &  & D^{L}\omega\left(Y|x\right)+e^{\alpha\dot{\beta}}y_{\alpha}\bar{\partial}_{\dot{\beta}}\Pi^{-}\omega\left(Y|x\right)+e^{\alpha\dot{\beta}}\partial_{\alpha}\bar{y}_{\dot{\beta}}\Pi^{+}\omega\left(Y|x\right)=\nonumber \\
 &  & =\dfrac{i}{4}\bar{H}^{\dot{\alpha}\dot{\beta}}\bar{\partial}_{\dot{\alpha}}\bar{\partial}_{\dot{\beta}}C\left(0,\bar{y}|x\right)+\dfrac{i}{4}H^{\alpha\beta}\partial_{\alpha}\partial_{\beta}C\left(y,0|x\right)+\nonumber \\
 &  & +\Biggl(\dfrac{i}{4}\bar{H}^{\dot{\alpha}\dot{\beta}}\bar{\partial}_{\dot{\alpha}}\bar{\partial}_{\dot{\beta}}\Pi^{+0}\dfrac{\bar{N}!\left(\bar{N}-2\right)!}{N+\bar{N}}\oint_{m=0}\dfrac{dm}{2\pi im}J\left(\dfrac{1}{m}y,\dfrac{1}{m}\bar{y}\right)+\nonumber \\
 &  & +\dfrac{i}{4}H^{\alpha\beta}y_{\alpha}y_{\beta}\dfrac{\bar{N}!\left(\bar{N}+1\right)!N!}{\left(N+3\right)!\left(N+\bar{N}+4\right)}\Pi^{+0}\oint_{m=0}\dfrac{dm}{2\pi im^{5}}T\left(\dfrac{1}{m}y,\dfrac{1}{m}\bar{y}\right)+h.c.\Biggr)|_{b=f=0},\label{OffShTh_flat_1}
\end{eqnarray}
\begin{eqnarray}
 &  & D^{L}C\left(Y|x\right)+ie^{\alpha\dot{\beta}}\partial_{\alpha}\bar{\partial}_{\dot{\beta}}C\left(Y|x\right)=-ie^{\alpha\dot{\beta}}y_{\alpha}\bar{y}_{\dot{\beta}}\dfrac{1}{\left(N+1\right)\left(N+2\right)}J-\nonumber \\
 &  & -\Biggl(e^{\alpha\dot{\beta}}y_{\alpha}\bar{\partial}_{\dot{\beta}}\frac{1}{\left(N+1\right)\left(N+2\right)\left(N-\bar{N}+2\right)}\Pi^{+0}\oint_{m=0}\dfrac{dm}{2\pi im^{3}}J\left(\dfrac{1}{m}y,m\bar{y}\right)+\nonumber \\
 &  & +ie^{\alpha\dot{\beta}}y_{\alpha}\bar{y}_{\dot{\beta}}\frac{1}{\left(N+1\right)\left(N+2\right)\left(N-\bar{N}\right)}\Pi^{+}\oint_{m=0}\dfrac{dm}{2\pi im}J\left(\dfrac{1}{m}y,m\bar{y}\right)+h.c.\Biggr)|_{b=f=0}.\label{OffShTh_flat_2}
\end{eqnarray}
As usual, all coefficients before HS sources get fixed by consistency
requirement \eqref{unf_consist}. Contour integrals in \eqref{OffShTh_flat_1}-\eqref{OffShTh_flat_2}
ensure that only a current of spin-$s$ (determined by power of $m$)
sources gauge field of spin-$s$ (determined by powers of $y$ and
$\bar{y}$): only when their spins coincide a pole arises. Coupling
of a current of some spin to a gauge field of another spin is forbidden
by consistency. Because new master-fields $J$ and $T$, introduced
to build off-shell completion, are 0-forms, it means, according to
\eqref{unf_gauge_transf}, that the gauge symmetries did not changed
under this completion, preserving their correct form. An on-shell
reduction of the system \eqref{OffShTh_flat_1}-\eqref{OffShTh_flat_2},
\eqref{T_eq}, \eqref{J_eq} is trivially achieved by putting $J=T=0$.

\section{Two-point functions\label{Sec4}}

One of the advantage of the proposed approach to building an off-shell
extension in comparison with the 'traceful' one is that introducing
auxiliary HS sources allows the direct looking for the partition function
of the corresponding quantum theory. The idea is that one can treat
a proper sector of the off-shell unfolded equations as the functional
Schwinger\textendash Dyson equations that determine the partition
function. In standard QFT if the classical equations of motion for
fields $\left\{ \varphi_{n}\right\} $ are
\begin{equation}
\dfrac{\delta S}{\delta\varphi_{n}}\left[\varphi_{k}\left(x\right)\right]=0,\label{class_EoM}
\end{equation}
then corresponding Schwinger\textendash Dyson equations for the partition
function $Z\left[J\right]=\int\mathcal{D}\varphi_{n}\mathrm{e}^{iS\left[\varphi\right]+iJ_{m}\varphi^{m}}$
are
\begin{equation}
\dfrac{\delta S}{\delta\varphi_{n}}\left[-i\dfrac{\delta}{\delta J_{k}\left(x\right)}\right]Z=-J_{n}\left(x\right)Z.\label{SchDys_eq}
\end{equation}
Introducing generating functional of connected correlators $W=-i\log Z$
and considering a free theory with linear e.o.m. one can reformulate
\eqref{SchDys_eq} as
\begin{equation}
\dfrac{\delta S_{free}}{\delta\varphi_{n}}\left[\dfrac{\delta W}{\delta J_{k}\left(x\right)}\right]=-J_{n}\left(x\right).\label{W_eq}
\end{equation}
(For construction of Schwinger\textendash Dyson equations in a non-Lagrangian
case see \cite{Lyakh}.) A transition from \eqref{class_EoM} to \eqref{W_eq}
is similar to the off-shell extension we considered.

Let us illustrate this scheme by evaluating two-point functions of
free HS fields from the off-shell system we found. One can solve corresponding
equations from \eqref{OffShTh_flat_1}-\eqref{OffShTh_flat_2} for
primary HS fields in terms of $J$ and $T$ and then treat them as
derivatives of $W$ with respect to the corresponding sources. In
Cartesian coordinates \eqref{Cartes_coord} equations of the form
\begin{eqnarray}
 &  & \left(D^{L}+ie^{\alpha\dot{\beta}}\partial_{\alpha}\bar{\partial}_{\dot{\beta}}\right)C\left(Y|x\right)=F\left(Y|x\right),\label{eq1}\\
 &  & \left(D^{L}+e^{\alpha\dot{\beta}}y_{\alpha}\bar{\partial}_{\dot{\beta}}\Pi^{-}+e^{\alpha\dot{\beta}}\partial_{\alpha}\bar{y}_{\dot{\beta}}\Pi^{+}\right)\omega\left(Y|x\right)=G\left(Y|x\right)\label{eq2}
\end{eqnarray}
for 0-form $C$ and 1-form $\omega=e^{\alpha\dot{\alpha}}\omega_{\alpha\dot{\alpha}}$
sourced by 1-form $F=e^{\alpha\dot{\alpha}}F_{\alpha\dot{\alpha}}$
and 2-form $G=\dfrac{1}{2}H^{\alpha\beta}G_{\alpha\beta}+\dfrac{1}{2}\bar{H}^{\dot{\alpha}\dot{\beta}}\bar{G}_{\dot{\alpha}\dot{\beta}}$
can be solved in the following manner
\begin{eqnarray}
 &  & C\left(Y|x\right)=-\dfrac{i}{2\left(2\pi\right)^{8}}\int\mathrm{d}^{4}p\int\mathrm{d}^{4}z\frac{\mathrm{e}^{ip\left(x-z\right)}}{p^{2}}\left(p_{\alpha\dot{\alpha}}-\partial_{\alpha}\bar{\partial}_{\dot{\alpha}}\right)F^{\alpha\dot{\alpha}}\left(Y|z\right),\\
 &  & \omega_{\alpha\dot{\alpha}}\left(Y|x\right)=-\dfrac{1}{2\left(2\pi\right)^{8}}\int\mathrm{d}^{4}p\int\mathrm{d}^{4}z\frac{\mathrm{e}^{ip\left(x-z\right)}}{p^{2}}\Bigl[ip^{\gamma}\text{}_{\dot{\alpha}}G_{\alpha\gamma}\left(Y|z\right)+ip_{\alpha}\text{}^{\dot{\gamma}}\bar{G}_{\dot{\alpha}\dot{\gamma}}\left(Y|z\right)-\nonumber \\
 &  & -\bar{y}^{\dot{\gamma}}\partial_{\alpha}\Pi^{+}\bar{G}_{\dot{\alpha}\dot{\gamma}}\left(Y|z\right)-y^{\gamma}\text{\ensuremath{\bar{\partial}}}_{\dot{\alpha}}\Pi^{-}G_{\alpha\gamma}\left(Y|z\right)-\text{\ensuremath{\bar{y}}}_{\dot{\alpha}}\partial^{\gamma}\Pi^{+}G_{\alpha\gamma}\left(Y|z\right)-y_{\alpha}\bar{\partial}^{\dot{\gamma}}\Pi^{-}\bar{G}_{\dot{\alpha}\dot{\gamma}}\left(Y|z\right)\Bigr].
\end{eqnarray}
In checking that these formulas indeed solve \eqref{eq1}-\eqref{eq2}
one has to use consistency conditions \eqref{unf_consist} for $F$
and $G$.

Then from \eqref{OffShTh_flat_1} and \eqref{J_eq} one finds an expression
for the (traceless components of) Fronsdal spin-$s>1$ field $\phi_{\alpha\left(s\right),\dot{\alpha}\left(s\right)}=\left(\left(s-1\right)!\right)^{2}\omega_{\underline{n},\alpha\left(s-1\right),\dot{\alpha}\left(s-1\right)}e^{\underline{n}}{}_{\alpha\dot{\alpha}}$
as a function of currents
\begin{eqnarray}
\phi_{\alpha\left(s\right),\dot{\alpha}\left(s\right)}\left(x\right) & = & \dfrac{i\left(\left(s-1\right)!\right)^{2}}{4\left(2\pi\right)^{8}\left(2s\right)!}\int\mathrm{d}^{4}p\int\mathrm{d}^{4}z\frac{\mathrm{e}^{ip\left(x-z\right)}}{p^{2}}\biggl(\left(1+s+s^{2}\right)J_{\alpha\left(s\right),\dot{\alpha}\left(s\right)}\left(z\right)-\nonumber \\
 &  & -s^{2}\frac{p_{\alpha\dot{\alpha}}p^{\beta\dot{\beta}}}{p^{2}}J_{\beta\alpha\left(s-1\right),\dot{\beta}\dot{\alpha}\left(s-1\right)}\left(z\right)\biggr).\label{fi_from_J}
\end{eqnarray}
The second $p$-proportional term in fact can be chosen arbitrarily
due to the gauge symmetry \eqref{unf_gauge_transf}, which for $\phi_{\alpha\left(s\right),\dot{\alpha}\left(s\right)}$
has the form
\begin{equation}
\delta\phi_{\alpha\left(s\right),\dot{\alpha}\left(s\right)}\left(x\right)=i\int\mathrm{d}^{4}p\mathrm{e}^{ipx}p_{\alpha\dot{\alpha}}\tilde{\epsilon}_{\alpha\left(s-1\right),\dot{\alpha}\left(s-1\right)}\left(p\right)
\end{equation}
for arbitrary $\tilde{\epsilon}_{\alpha\left(s-1\right),\dot{\alpha}\left(s-1\right)}\left(p\right)$.
So for simplicity we gauge away the second term in \eqref{fi_from_J}
(impose the Feynman gauge)
\begin{equation}
\phi_{\alpha\left(s\right),\dot{\alpha}\left(s\right)}\left(x\right)=\dfrac{i\left(\left(s-1\right)!\right)^{2}}{4\left(2\pi\right)^{8}\left(2s\right)!}\left(1+s+s^{2}\right)\int\mathrm{d}^{4}p\int\mathrm{d}^{4}z\frac{\mathrm{e}^{ip\left(x-z\right)}}{p^{2}}J_{\alpha\left(s\right),\dot{\alpha}\left(s\right)}\left(z\right).\label{fi_J_Feyn_gauge}
\end{equation}
Analogously one can find expressions for electromagnetic and scalar
fields, using \eqref{OffShTh_flat_2}.

Now treating \eqref{fi_J_Feyn_gauge} in the spirit of \eqref{W_eq}
as the $J$-derivative of $W$
\begin{equation}
\phi_{\alpha\left(s\right),\dot{\alpha}\left(s\right)}\left(x\right)=-\dfrac{\delta W}{\delta J^{\alpha\left(s\right),\dot{\alpha}\left(s\right)}}\left(x\right)
\end{equation}
and considering that connected correlators are given by $G^{\left(c\right)}\left(x_{1},...x_{n}\right)=\dfrac{\delta}{i\delta J\left(x_{1}\right)}...\dfrac{\delta}{i\delta J\left(x_{n}\right)}iW|_{J=0}$
one finds two-point correlation functions for traceless HS fields
\begin{equation}
\left\langle \phi_{\alpha\left(s\right),\dot{\alpha}\left(s\right)}\left(x_{1}\right)\phi_{\beta\left(s\right),\dot{\beta}\left(s\right)}\left(x_{2}\right)\right\rangle =k_{s}\int\mathrm{d}^{4}p\frac{\mathrm{e}^{ip\left(x_{1}-x_{2}\right)}}{p^{2}}\left(\epsilon_{\alpha\beta}\right)^{s}\left(\epsilon_{\dot{\alpha}\dot{\beta}}\right)^{s},
\end{equation}
with $k_{s}$ being $s$-dependent constants that can be properly
normalized by rescaling of $J$.

\section{Conclusion\label{Conclusion}}

In this paper we found an off-shell completion for unfolded system
of free HS fields in $4d$ flat spacetime in spinorial formalism.
In order to do this we derived an unfolded description of HS currents.
These currents, when coupled to on-shell unfolded HS equations, remove
differential constraints on HS fields and get an interpretation of
descendants of these fields. As byproduct we found an unfolded spinorial
description of unconstrained traceless Lorentz tensor field of arbitrary
rank, whose reduction, in particular, corresponds to conformal HS
currents, which may help in the study of this subject. We also showed
that such an off-shell unfolded system can be reinterpreted as Schwinger\textendash Dyson
system, and restored two-point functions of HS fields this way.

The next issue to be studied is the deformation of the presented system
to $AdS$ space, where the full-fledged nonlinear HS theory lives.
One may hope that at the end of the day this will allow one to build
an off-shell completion for the full $4d$ Vasiliev theory, making
possible the systematic study of HS action principle problem as well
as the quantization issue.

\section*{Acknowledgments}

The author is grateful to M.A. Grigoriev, D.S. Ponomarev and M.A.
Vasiliev for valuable discussions and useful remarks. The author also
thanks the Erwin Schr$\ddot{\textrm{o}}$dinger Institute in Vienna
for hospitality during the program ``Higher Spins and Holography''
where the part of the work was done. The research was supported by
the Russian Science Foundation grant 18-12-00507.

\end{document}